\documentclass[12pt]{iopart}

\usepackage{graphicx}
\usepackage{amssymb}
\usepackage{citesort}
\DeclareRobustCommand{\<}[1]{\hspace{-0.11111em}#1\hspace{-0.11111em}}
\DeclareRobustCommand{\rtrim}[1]{#1\hspace{-0.11111em}}

\usepackage{amstext}
\DeclareRobustCommand{\lvert}{|}
\DeclareRobustCommand{\rvert}{|}

\DeclareRobustCommand{\grp}[1]{\mathrm{#1}}
\DeclareRobustCommand{\grpun}{\grp{U}(n)}
\DeclareRobustCommand{\grputhree}{\grp{U}(3)}
\DeclareRobustCommand{\grputwo}{\grp{U}(2)}
\DeclareRobustCommand{\grpsothree}{\grp{SO}(3)}
\DeclareRobustCommand{\grpsotwo}{\grp{SO}(2)}

\DeclareRobustCommand{\Nhat}{\hat{N}}

\DeclareRobustCommand{\lhat}{\hat{l}}
\DeclareRobustCommand{\Dhat}{\hat{D}}

\DeclareRobustCommand{\That}{\hat{T}}
\DeclareRobustCommand{\Qhat}{\hat{Q}}
\DeclareRobustCommand{\What}{\hat{W}}

\DeclareRobustCommand{\falling}[2]{{{#1}^{\underline{#2}}}}
\DeclareRobustCommand{\fallingsmash}[2]{{{#1}{}^{\smash{\underline{#2}}}}}
\DeclareRobustCommand{\alphaindex}[4]{{\alpha_{{#1},r_{#2}{#3}}^{#4}}}

\begin{document}

\title{Application of the coherent state formalism to multiply excited states}

\author{M~A~Caprio}

\address{Center for Theoretical Physics, Sloane Physics Laboratory, 
Yale University, New Haven, Connecticut 06520-8120, USA}

\begin{abstract}
A general expression is obtained for the matrix element of an $m$-body
operator between coherent states constructed from multiple orthogonal
coherent boson species.  This allows the coherent state formalism to
be applied to states possessing an arbitrarily large number of
intrinsic excitation quanta.  For illustration, the formalism is
applied to the two-dimensional vibron model [$\grputhree$ model], to
calculate the energies of all excited states in the large-$N$ limit.
\end{abstract}

\pacs{03.65.Fd, 33.20.Tp}

\submitto{\JPA}


\section{Introduction}

Methods based upon coherent states~\cite{zhang1990:coherent} have
proved to be widely useful in investigating the ground state
properties of algebraic models.  In algebraic models, the Hamiltonian
and all physical operators are constructed from the elements of a Lie
algebra, usually from a bosonic realization of $\grpun$.  Such models
have been applied extensively to the spectroscopy of many-body
systems, including nuclei~\cite{iachello1987:ibm} and
molecules~\cite{iachello1995:vibron}.  The coherent states of an
algebraic model are obtained by repeated action of a general linear
combination of boson creation operators on the vacuum state.  As
variational trial states, the coherent states allow the estimation of
the ground state energies and properties, yielding results which
become exact in the infinite boson number
limit~\cite{gilmore1975:multilevel-coherent,gilmore1978:coherent}.
They are also essential in defining the classical limit for the
model~\cite{gilmore1978:coherent}, providing the geometric coordinates
or dynamical variables of the model.

Coherent states may also be used to study the intrinsic excitation
modes of an algebraic
system~\cite{bohr1982:ibm-coherent-excited-168er,bijker1982:ibm-coherent-excited,leviatan1985:ibm-o5-coherent,leviatan1987:ibm-intrinsic,kuyucak1988:ibm-expansion,leviatan1988:vibron-intrinsic,kuyucak1989:ibm2-expansion,leviatan1990:ibm2-modes,leviatan1991:ibm2-modes-nonaxial,garciaramos1998:ibm-two-phonon},
through the construction of coherent states orthogonal to the ground
state.  However, application of this method has generally been limited
to excited states involving only one intrinsic excitation quantum, or
at most two~\cite{garciaramos1998:ibm-two-phonon}, due to the
complexity of calculating the necessary matrix elements.  In the
present work, a general expression is obtained for the matrix element
of an $m$-body operator between coherent states constructed from
multiple orthogonal linear combinations of boson creation operators.
This allows the coherent state formalism to be applied to states
possessing an arbitrarily large number of intrinsic excitation quanta.

The necessary definitions for the basic ground state (condensate)
coherent state are first presented, and the results for expectation
values with respect to this state are reviewed
(section~\ref{seccondensate}).  The general result for excited
coherent states is then established (section~\ref{secmulti}).  As a
simple illustration, calculations are carried out for excited states
in the $\grpsothree$ dynamical symmetry limit of the molecular
two-dimensional vibron model~\cite{iachello1996:vibron-2dim}
(section~\ref{secvibron}).

\section{Condensate coherent state}
\label{seccondensate}

An algebraic model based upon a bosonic realization of $\grpun$ is
obtained by defining $n$ bosonic creation operators $b_1^\dagger$,
$b_2^\dagger$,$\ldots$, $b_n^\dagger$ obeying the canonical
commutation relations $[b_i,b_j]\<=0$,
$[b_i,b_j^\dagger]\<=\delta_{i,j}$, and
$[b_i^\dagger,b_j^\dagger]\<=0$.  The set of all possible bilinears of
a creation and an annihilation operator then forms a basis
$b_1^\dagger b_1$, $b_1^\dagger b_2$,~$\ldots$, $b_n^\dagger b_n$ for
the algebra.  The physical operators of the model, such as the
Hamiltonian, the angular momentum, and transition operators, are
constructed as polynomials in the $b_i^\dagger b_j$.  These operators
act on the states of the Fock space created by $b_1^\dagger$,
$b_2^\dagger$,~$\ldots$, $b_n^\dagger$.

The condensate coherent state is defined in terms of the ``condensate
boson'' creation operator, which is a general linear combination
\begin{equation}
\label{eqnBc}
B_c^\dagger \equiv \alpha_1 b_1^\dagger+\alpha_2 b_2^\dagger+\cdots+\alpha_n b_n^\dagger,
\end{equation}
with complex coefficients satisfying the normalization convention
$\sum_{i=1}^n \alpha_i^*
\alpha_i\<=1$.  The normalized condensate coherent state is then
\begin{equation}
\label{eqncoherentc}
|N;\alpha_1,\ldots,\alpha_n\rangle 
\equiv \frac{1}{\sqrt{N!}}(B_c^\dagger) ^N |0\rangle.
\end{equation}
This state is an eigenstate of the total number operator $\Nhat\<\equiv\sum_{i=1}^n b_i^\dagger
b_i$.

The expectation value of a one-body or two-body operator with respect
to the condensate~(\ref{eqncoherentc}) was deduced by Van~Isacker and
Chen~\cite{vanisacker1981:ibm-triax}, using arguments based upon
formal differentiation.  Here let us derive an explicit result for the
expectation value of an arbitrary $m$-body operator, since the results
of section~\ref{secmulti} can be obtained as a natural extension.
First, for an annihilation operator $b_r$, note that
$[b_r,B_c^\dagger]\<=\alpha_r$, from which the relation
\begin{equation}
\label{eqncommutatorc}
[b_r,(B_c^\dagger)^N]=N\alpha_r(B_c^\dagger)^{N-1}
\end{equation}
follows by the product rule for commutators.
The action of $b_r$ on the condensate is thus
\begin{equation}
\label{eqnactionc}
b_r|N;\alpha_1,\ldots,\alpha_n\rangle=\sqrt{N} \alpha_r |N-1;\alpha_1,\ldots,\alpha_n\rangle.
\end{equation}
Repeated application yields the expectation value of an arbitrary
$m$-body operator,
\begin{equation}
\label{eqnmec}
\langle N;\alpha_1,\ldots,\alpha_n|
\Bigl(\prod_{i=1}^m b^\dagger_{r_i'}\Bigr)
\Bigl(\prod_{i=1}^m b_{r_i}\Bigr)
|N;\alpha_1,\ldots,\alpha_n\rangle
=
N^{\underline{m}}
\prod_{i=1}^m \alpha^*_{r_i'}\alpha_{r_i},
\end{equation}
where the underlined superscript indicates the falling
factorial~\cite{graham1994:concrete}, 
defined by $m^{\underline{n}}\<\equiv
m(m-1)\cdots(m-n+1)$.

\section{General coherent state}
\label{secmulti}

For the study of excited states, it is necessary to consider coherent
states which are orthogonal to the condensate state.  These are
constructed using multiple different coherent boson species $B_s^\dag$
($s\<=1,\ldots,S$), defined as linear combinations
\begin{equation}
\label{eqnBmulti}
B_s^\dagger \equiv \alpha_{s,1} b_1^\dagger+\alpha_{s,2} b_2^\dagger+\cdots+\alpha_{s,n} b_n^\dagger
\end{equation}
of the basic boson creation operators.  
(The procedure for obtaining values of the
coefficients $\alpha_{s,i}$ appropriate to a given model is discussed
in, \textit{e.g.}, ref.~\cite{leviatan1988:vibron-intrinsic}.)  The coefficients
$\alpha_{s,i}$ are chosen to obey the orthonormalization convention
$\sum_{i=1}^n \alpha_{s',i}^*
\alpha_{s,i}\<=\delta_{s',s}$.  Consequently, the coherent bosons satisfy
canonical commutation relations $[B_{s'},B_{s}]\<=0$,
$[B_{s'},B_{s}^\dagger]\<=\delta_{s',s}$, and
$[B_{s'}^\dagger,B_{s}^\dagger]\<=0$.  The different $B_s^\dag$
represent the ground state condensate boson and one or more orthogonal
excitation modes.    The
normalized multi-species coherent state is
\begin{equation}
\label{eqncoherentmulti}
\lvert N_1\cdots N_S\rangle 
\equiv \Bigl( \prod_{s=1}^S
\frac{1}{\sqrt{N_s!}}(B_s^\dagger)^{N_s}\Bigr) |0\rangle.
\end{equation}
The coherent state is an eigenstate of the total number operator, of
eigenvalue $N\<=\sum_{s=1}^SN_s$.  States of different
coherent boson occupation numbers $N_1$,~$\ldots$, $N_S$ are orthogonal.

The matrix element of an arbitrary $m$-body operator with respect to
two multi-species coherent states can now be deduced following the
approach of section~\ref{seccondensate}.  The commutation relations
generalize to $[b_r,B_s^\dagger]\<=\alpha_{s,r}$ and
\begin{equation}
\label{eqncommutatormulti}
[b_r,\prod_{s=1}^S(B_s^\dagger)^{N_s}]
=\sum_{t=1}^S N_t \alpha_{t,r}\Bigl[ \prod_{s=1}^S(B_s^\dagger)^{N_s-\delta_{s,t}} \Bigr].
\end{equation}
The action of $b_r$ on the multi-species coherent state is thus
\begin{equation}
\label{eqnactionmulti}
b_r|N_1\cdots N_S\rangle=\sum_{t=1}^S \sqrt{N_t} \alpha_{t,r} |(N_1-\delta_{1,t})\cdots(N_S-\delta_{S,t})\rangle.
\end{equation}
This is analogous to the result of equation~(\ref{eqnactionc}), but
with a separate term arising from the action of $b_r$ on each species
of coherent boson contributing to the coherent state.  It is useful to
define a counting function
$\nu(t_1,\ldots,t_m;s)\<\equiv\sum_{i=1}^m\delta_{t_i,s}$,
giving the number of the $t_i$ which are equal to $s$.
Then, for a product of annihilation operators acting on the coherent state,
\begin{equation}
\label{eqnprodactionmulti}
\eqalign{
\fl
\Bigl(\prod_{i=1}^m b_{r_i}\Bigr)|N_1\cdots N_S\rangle
=
\sum_{t_1,\ldots,t_m=1}^S \Bigl[\prod_{s=1}^S
\sqrt{N_s^{\underline{\nu(t_1,\ldots,t_m;s)}}} \Bigr]
\Bigl( \prod_{i=1}^m \alpha_{t_i,r_i} \Bigr)
\\
\times
\bigl|
\bigl(N_1-\nu(t_1,\ldots,t_m;1)\bigr)\cdots\bigl(N_S-\nu(t_1,\ldots,t_m;S)\bigr)\bigr\rangle.
}
\end{equation}
The matrix element of an arbitrary $m$-body operator
($m\<\geq1$) with respect to two arbitrary multi-species coherent states
is the inner product of two such expressions, 
\begin{equation}
\label{eqnmemulti}
\eqalign{
\fl
\langle N_1'\cdots N_S'\rvert \Bigl(\prod_{i=1}^m b_{r_i'}^\dagger\Bigr)
\Bigl(\prod_{i=1}^m b_{r_i}\Bigr)|N_1\cdots N_S\rangle
\\
\lo=
\sum_{{t_1',\ldots,t_m' \atop t_1,\ldots,t_m} =1}^S 
\Bigl[\prod_{s=1}^S
\delta_{N_s'-\nu_s',N_s-\nu_s}
\sqrt{N_s'^{\underline{\nu_s'}}N_s^{\underline{\nu_s}}} \Bigr]
\Bigl( \prod_{i=1}^m \alpha_{t_i',r_i'}^* \alpha_{t_i,r_i} \Bigr),
}
\end{equation}
where the abbreviations $\nu_s'\<\equiv\nu(t_1',\ldots,t_m';s)$ and
$\nu_s\<\equiv\nu(t_1,\ldots,t_m;s)$ have been used.

Three stages are involved in evaluating the matrix element of a
general operator: reexpression of the operator in terms of
normal-ordered $m$-body terms, evaluation of the matrix elements of
these by equation~(\ref{eqnmemulti}), and simplification of the
result.  For complicated operators or if many coherent boson species
are involved, these steps can most effectively be carried out though
computer-based symbolic manipulation.  A few useful special cases of
equation~(\ref{eqnmemulti}) are summarized in the appendix.

The multiple sum in equation~(\ref{eqnmemulti}) nominally contains
$S^{2m}$ terms. However, for fixed numerical values of the $N_s$ and $N_s'$,
many of the terms vanish identically due to the restriction on indices
imposed by the product of Kronecker delta symbols.  A summary of the number of
nonvanishing terms for various $S$ and $m$ is given in
table~\ref{tabterms}.  A given species $s$ of coherent boson is overannihilated
when $N_s-\nu(s)\<<0$ or $N_s'-\nu'(s)\<<0$, yielding a vanishing falling factorial in
equation~(\ref{eqnmemulti}).  Thus, additional terms vanish if the expression is
evaluated for a small value ($\rtrim<m$) of any of the $N_s$ or
$N_s'$, as typically occurs when the lowest-lying excited states are
considered.  If the multiple sum in
equation~(\ref{eqnmemulti}) is instead to be evaluated with the $N_s$ and $N_s'$
retained as variables, all terms involving the same product of
falling factorials may be collected.  This product is identical for
terms with the same values of all the $\nu(s)$ and $\nu'(s)$.  Since
$0\<\leq\nu(s)\<\leq m$ and $\sum_{s=1}^S\nu(s)\<=m$, and similarly
for $\nu'(s)$, the number of distinct terms after collection is the
square of the number of possible partitions of $m$ over $S$ bins.
\begin{table}
\caption{\label{tabterms}The number of terms in general contributing to the sum
in equation~(\ref{eqnmemulti}), not vanishing due to the delta
symbol constraint, for various values of $S$ and $m$.  In practice,
some of these terms may also vanish as a consequence of zero values
for the falling factorials or $\alpha_{s,i}$ coefficients.}
\begin{indented}\item[]
\begin{tabular}{@{}rrrrr}
\br
&\centre{4}{Contributing terms}
\vspace{-3pt}
\\
\ns
&\crule{4}\\
$S$ & $1$-body & $2$-body & $3$-body & $4$-body\\
\mr
 2 &   2 &   6 &  20 &  70\\
 3 &   3 &  15 &  93 & 639\\
 4 &   4 &  28 & 256 & 2716\\
 5 &   5 &  45 & 545 & 7885\\
\br
\end{tabular}
\end{indented}
\end{table}

Multi-species coherent states of the form~(\ref{eqncoherentmulti}) are
also encountered as the condensate states of systems involving
multiple constituents, each separately conserved.  An example from
nuclear physics is the proton-neutron interacting boson model
(IBM-2)~\cite{iachello1987:ibm}, in which proton pairs (created by
$s_{\pi,0}^{\dagger}$, $d_{\pi,-2}^{\dagger}$, $d_{\pi,-1}^{\dagger}$,
$d_{\pi,0}^{\dagger}$, $d_{\pi,+1}^{\dagger}$, $d_{\pi,+2}^{\dagger}$)
and neutron pairs (created by $s_{\nu,0}^{\dagger}$,
$d_{\nu,-2}^{\dagger}$, $d_{\nu,-1}^{\dagger}$, $d_{\nu,0}^{\dagger}$,
$d_{\nu,+1}^{\dagger}$, $d_{\nu,+2}^{\dagger}$) are separately
conserved.  The physical operators are constructed from the elements
of a Lie algebra $\grp{U}_1(n)\otimes\grp{U}_2(n)\otimes\cdots$, and
a condensate state with good boson number for each constituent is
constructed as
$\rtrim\propto(B_{c1}^\dag)^{N_{1}}(B_{c2}^\dag)^{N_{2}}\cdots\lvert 0
\rangle$.  Since the condensate bosons $B_{c\rho}^\dag$
($\rho\<=1,2,\ldots$) are constructed from disjoint sets of boson
operators, the expectation value of an $m$-body operator in general
factorizes into the product of simple expectation values of the
type~(\ref{eqnmec}) (\textit{e.g.},
ref.~\cite[(C1)]{ginocchio1992:ibm2-shapes}).  However, the general
result~(\ref{eqnmemulti}) for the multi-species coherent state matrix
element can provide the simplest framework for computer-based symbolic
evaluation~\cite{caprio2005:ibmpn2}.

\section{Intrinsic excitations of the two-dimensional vibron model}
\label{secvibron}

The two-dimensional vibron model~\cite{iachello1996:vibron-2dim} is
the $\grputhree$ algebraic model, describing a system containing a
dipole degree of freedom constrained to planar motion.  The basic
example of such a system is a triatomic linear bender molecule, but
the model is easily extended to more complex molecular systems.  The
$\grputhree$ algebra is realized in terms of the three bosonic
operators $\sigma^\dag$, $\tau_x^\dag$, and $\tau_y^\dag$, which satisfy
canonical commutation relations.  It is convenient to define circular
bosons $\tau_\pm^\dag\<\equiv\mp(\tau_x^\dag\pm
i\tau_y^\dag)/\sqrt{2}$~\cite{perez-bernal:PC,perez-bernal2005:vibron-2dim-nonrigidity,endnote-sign},
such that the operators $\sigma^\dag$, $\tau_+^\dag$, and $\tau_-^\dag$ carry
$0$, $+1$, and $-1$ units of two-dimensional angular momentum.
The physical operators include the angular
momentum operator $\lhat\<\equiv\tau_+^\dag\tau_+-\tau_-^\dag\tau_-$,
the dipole operators
$\Dhat_\pm\<\equiv\pm\sqrt{2}(\tau_\pm^\dag\sigma-\sigma^\dag\tau_\mp)$,
and the quadrupole operators
$\Qhat_\pm\<\equiv\sqrt{2}\tau_\pm^\dag\tau_\mp$.

\begin{figure}
\begin{indented}\item[]
\includegraphics*[width=0.8\hsize]{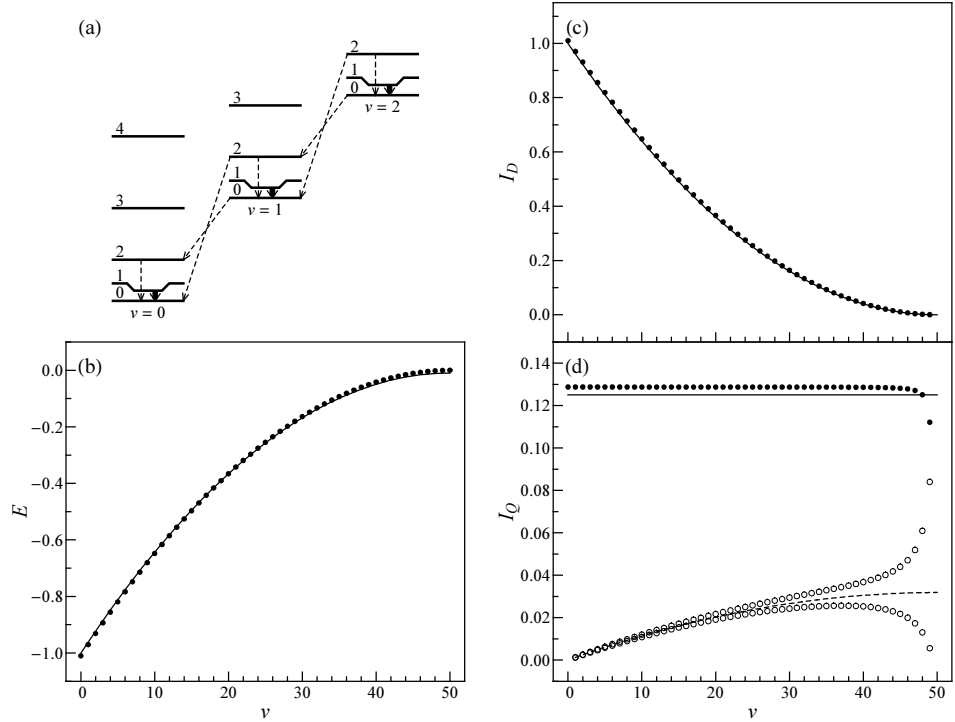}
\end{indented}
\caption{Comparison of 
excited coherent state estimates of observables~(curves) with exact
values~(circles) for the $\grpsothree$ dynamical symmetry of the
two-dimensional vibron model, with $N\<=100$.  (a)~Schematic
$\grpsothree$ level energy diagram, with arrows indicating the
dipole~(solid) and quadrupole~(dashed) transitions considered in
panels~(c,d).  (b)~Energies of all $\grpsothree$ representations.
(c)~Intensity of the $(l\<=1)\<\rightarrow(l\<=0)$ dipole transition
within each representation.  (d)~Intensity of the
$(l\<=2)\<\rightarrow(l\<=0)$ quadrupole transition within each
representation (solid curve, filled circles) and of the
$(l\<=2)\<\rightarrow(l\<=0)$ and $(l\<=0)\<\rightarrow(l\<=2)$
$\Delta v\<=-1$ transitions between representations (dashed curve,
open circles).  An $l^2$ energy splitting is included in panel~(a), to
provide visual separation of levels within an $\grpsothree$
representation.  All observables are plotted rescaled by $1/N^2$.  }
\label{figcombo}
\end{figure}

The $\grputhree$ algebra contains the subalgebra
chains~\cite{iachello1996:vibron-2dim}
\begin{equation}
\label{eqnchains}
\grputhree\supset\cases{
\grputwo\supset\grpsotwo  \qquad &\\
\grpsothree\supset\grpsotwo. &\\
}
\end{equation}
The dynamical symmetry associated with the $\grputwo$ chain yields
spectra matching those of the cylindrical oscillator (P\"oschl-Teller
potential), while the dynamical symmetry associated with the
$\grpsothree$ chain yields spectra like those of the displaced
cylindrical oscillator (Morse potential).  The $\grpsothree$ limit is
used in the following illustrations, as the less trivial case.
The $\grpsothree$ subalgebra is spanned by $\lbrace
\Dhat_+,\Dhat_-,\lhat\rbrace$ and has quadratic Casimir 
operator $\What^2\<\equiv(\Dhat_+\Dhat_-+\Dhat_-\Dhat_+)/2+\lhat^2$.
The $\grpsotwo$ subalgebra is simply the two-dimensional angular
momentum algebra, containing $\lhat$.  The simplest Hamiltonian with
$\grpsothree$ dynamical symmetry is $H\<=-\What$, which has
eigenvalues
\begin{equation}
\label{vibronenergyso3}
E(N,v,l)=-N(N+1)+4v[(N+1/2)-v],
\end{equation}
with $v\<=0,1,\ldots,\lfloor N/2 \rfloor$ and
$l\<=-(N-2v),-(N-2v)+1,\ldots,+(N-2v)$ [figure~\ref{figcombo}(a)].%

The condensate boson and an orthogonal excitation boson for the
two-dimensional vibron model may be defined as
\begin{equation}
B_c^\dag(r)\equiv\frac{1}{\sqrt{1+r^2}}(\sigma^\dag+r\tau_x^\dag) 
\qquad
B_x^\dag(r)\equiv\frac{1}{\sqrt{1+r^2}}(-r\sigma^\dag+\tau_x^\dag).
\end{equation}
(See ref.~\cite{leviatan1988:vibron-intrinsic} for further discussion
of the choice of boson operators in the vibron model.)
The general excited coherent state is
\begin{equation}
\label{eqnvibroncoherent}
\lvert NN_x;r\rangle\equiv \frac{1}{\sqrt{(N-N_x)!N_x!}}[B_c^\dag(r)]^{N-N_x} [B_x^\dag(r)]^{N_x} \lvert0\rangle.
\end{equation}
The expectation value of the $\grpsothree$ Casimir
operator with respect to an arbitrary excited coherent state is
evaluated using equation~(\ref{eqnmetwospeciestwobody}), yielding
\begin{equation}
\label{eqnvibronevW}
\eqalign{
\fl\langle NN_x;r\rvert\, \What^2 \,\lvert  NN_x;r\rangle 
\\
\lo=2[N+N_x(N-N_x)]+\frac{4}{(1+r^2)^2}[N(N-1)-6N_x(N-N_x)]r^2.
}
\end{equation}
The equilibrium value of $r$ is found by minimization of the
variational energy $\langle N0;r\rvert H
\lvert N0;r\rangle$, giving $r\<=1$.  With this value of $r$, 
the excited coherent state energies are
\begin{equation}
\label{vibronevWso3}
E(N,N_x)=-N(N+1)+4N_x[N-N_x].
\end{equation}
If the intrinsic excitation number $N_x$ is identified with the
$\grpsothree$ quantum number $v$, comparison of
equations~(\ref{vibronenergyso3}) and~(\ref{vibronevWso3}) shows that
these expressions differ only by a term of order $1/N$.  Thus, to
leading order in $1/N$, the coherent state estimate reproduces the
excitation energies for all excited states, as illustrated in
figure~\ref{figcombo}(b).

Transition strengths may be estimated
using coherent states, from the squared matrix element of the
transition operator, 
$I\<\approx\langle N N_x'\rvert \That \lvert N N_x
\rangle^2$.  The coherent state $\lvert N N_x \rangle$ is not an
angular momentum eigenstate, so the resulting estimate for the
transition strength between two intrinsic excitations is effectively
averaged over the many states of different angular momenta
constituting that excitation.  This coherent state estimate thus
cannot be expected to provide the exact transition intensity between
any two particular angular momentum eigenstates.  It does, however,
indicate the general magnitude of the transition strengths and the
overall dependence on excitation quantum number, and it can be
quantitatively accurate if the angular momentum dependence of
transition strengths is weak.  (Alternatively, angular momentum
eigenstates may be projected from the coherent
states~\cite{peierls1957:shell-model-projection,leviatan1987:ibm-intrinsic,leviatan1988:vibron-intrinsic},
but this requires additional machinery beyond simple evaluation of an
$m$-body operator matrix element.)

Dipole infrared transitions in the two-dimensional vibron model are,
to leading order, induced by the operators $\Dhat_\pm$, and quadrupole
Raman transitions are induced by the operators
$\Qhat_\pm$~\cite{iachello1995:vibron,iachello1996:vibron-2dim}.  The
strengths of transitions within an intrinsic excitation are estimated
from the expectation values
\begin{equation}
\label{eqnvibronme0}
\eqalign{
\langle NN_x;r\rvert\, \Dhat_\pm \,\lvert  NN_x;r\rangle 
&=-2(N-2N_x)\frac{r}{1+r^2}\\
\langle NN_x;r\rvert\, \Qhat_\pm \,\lvert  NN_x;r\rangle 
&=-\frac{1}{\sqrt{2}}\frac{N_x+(N-N_x)r^2}{1+r^2},
}
\end{equation}
and those between succesive intrinsic excitations from
\begin{equation}
\label{eqnvibronme1}
\eqalign{
\langle N(N_x-1);r\rvert\, \Dhat_\pm \,\lvert  NN_x;r\rangle 
&=-\sqrt{(N-N_x+1)N_x}\frac{1-r^2}{1+r^2}\\
\langle N(N_x-1);r\rvert\, \Qhat_\pm \,\lvert  NN_x;r\rangle 
&=-\frac{1}{\sqrt{2}}\sqrt{(N-N_x+1)N_x}\frac{r}{1+r^2}.
}
\end{equation}
The coherent state estimates of transition intensities for the
$\grpsothree$ dynamical symmetry are obtained from these equations
with $r\<=1$.  The estimates are plotted for $N\<=100$ in
figure~\ref{figcombo}(c,d), together with exact values obtained by
numerical diagonalization, as functions of the excitation quantum
number $v$ (or $N_x$).  The coherent state estimate for dipole
transitions closely reproduces the strength of the angluar momentum
$1\<\rightarrow0$ transition within an $\grpsothree$ representation
[figure~\ref{figcombo}(c)].  Dipole transitions between different
representations are forbidden, and the coherent state estimate indeed
vanishes.  Quadrupole transition strengths exhibit greater angular
momentum dependence within a representation, and the coherent state
estimate is consequently less accurate.  The strengths of quadrupole
transitions involving the low angular momentum members of the
representations are reproduced to within $\rtrim\sim5\%$, except at
the highest intrinsic excitation quantum numbers
[figure~\ref{figcombo}(d)].  Note that the angular momentum
$2\<\rightarrow0$ and $0\<\rightarrow2$ transitions between two
representations ($\Delta v\<=\pm1$) differ in strength, and the
coherent state estimate consistently behaves as their average.

\section{Conclusion}
\label{secconcl}

The present results serve as a basis for application of the coherent
state formalism to states with an arbitrary number of intrinsic
excitation quanta.  This process yields estimates of eigenvalues and
operator matrix elements for excited states valid to leading order in
$1/N$.  The illustration provided was to a dynamical symmetry limit of
a simple model, but the coherent state analysis will likely be most
useful when applied to transitional Hamiltonians, for which analytic
results are not otherwise available.

The coherent state formalism has in the past provided not only a
quantitative calculational tool but, perhaps more importantly, a
method for obtaining qualitative understanding of the equilibrium
properties and fundamental modes of a system.  Most, if not all, of
the raw numerical results of the coherent state formalism can also be
obtained by numerical diagonalization.  It is thus the latter,
interpretational aspects of the coherent state formalism, and the
explicit analytic forms obtained for the parameter dependences of
observables, which have proved most useful.  The present results for
multiply excited states thus might most productively be used in
investigating the general nature of the evolution of a system's
properties with excitation energy.

\ack

Discussions with F.~Iachello, A.~Leviatan, and F.~P\'erez-Bernal are
gratefully acknowledged.  This work was supported by the US DOE under
grant DE-FG02-91ER-40608.

%
%
%
%

\appendix
\def\thesection{Appendix}   
\section{Special cases of the general matrix element}

Some useful special cases of the general matrix
element~(\ref{eqnmemulti}) are given here.  If the coherent states
involve only two species of coherent boson ($S\<=2$), then the 1-body
and two-body operator matrix elements are
\begin{equation}
\label{eqnmetwospeciesonebody}
\eqalign{
\langle (N_1-1)(N_2+1) \rvert\, b_{r'}^\dagger b_r \,\lvert N_1 N_2 \rangle
&=\sqrt{N_1(N_2+1)}\, \alpha_{2,r'}^*\alpha_{1,r}
\\
\langle N_1N_2 \rvert\, b_{r'}^\dagger b_r \,\lvert N_1 N_2 \rangle
&=N_1\,\alpha_{1,r'}^*\alpha_{1,r} + N_2\,\alpha_{2,r'}^*\alpha_{2,r} 
\\
\langle (N_1+1)(N_2-1) \rvert\, b_{r'}^\dagger b_r \,\lvert N_1 N_2 \rangle
&=\sqrt{(N_1+1)N_2}\, \alpha_{1,r'}^*\alpha_{2,r}
}
\end{equation}
and
\begin{equation}
\label{eqnmetwospeciestwobody}
\eqalign{
\fl
\langle (N_1-2)(N_2+2) \rvert\, b_{r_2'}^\dagger b_{r_1'}^\dagger b_{r_1} b_{r_2} \,\lvert N_1 N_2 \rangle
=
\sqrt{\fallingsmash{N_1}{2}\fallingsmash{(N_2+2)}{2}}\,
\alphaindex{2}{2}{'}{*}\alphaindex{2}{1}{'}{*}\alphaindex{1}{1}{}{}\alphaindex{1}{2}{}{}
\\
\fl\langle (N_1-1)(N_2+1) \rvert\, b_{r_2'}^\dagger b_{r_1'}^\dagger b_{r_1} b_{r_2} \,\lvert N_1 N_2 \rangle
\\
\lo=
(N_1-1)\sqrt{N_1(N_2+1)}\,
\bigl(\alphaindex{1}{2}{'}{*}\alphaindex{2}{1}{'}{*}+\alphaindex{2}{2}{'}{*}\alphaindex{1}{1}{'}{*}\bigr)
\alphaindex{1}{1}{}{}\alphaindex{1}{2}{}{}\\
+N_2\sqrt{N_1(N_2+1)}\,\alphaindex{2}{2}{'}{*}\alphaindex{2}{1}{'}{*}
\bigl(\alphaindex{2}{1}{}{}\alphaindex{1}{2}{}{}+\alphaindex{1}{1}{}{}\alphaindex{2}{2}{}{}\bigr)
\\
\fl\langle N_1N_2 \rvert\, b_{r_2'}^\dagger b_{r_1'}^\dagger b_{r_1} b_{r_2} \,\lvert N_1 N_2 \rangle
=
\falling{N_1}{2}\,\alphaindex{1}{2}{'}{*}\alphaindex{1}{1}{'}{*}\alphaindex{1}{1}{}{}\alphaindex{1}{2}{}{}
+\falling{N_2}{2}\,\alphaindex{2}{2}{'}{*}\alphaindex{2}{1}{'}{*}\alphaindex{2}{1}{}{}\alphaindex{2}{2}{}{}
\\
+ N_1N_2\,\bigl(\alphaindex{1}{2}{'}{*}\alphaindex{2}{1}{'}{*}+\alphaindex{2}{2}{'}{*}\alphaindex{1}{1}{'}{*}\bigr)
\bigl(\alphaindex{1}{1}{}{}\alphaindex{2}{2}{}{}+\alphaindex{2}{1}{}{}\alphaindex{1}{2}{}{}\bigr)
\\
\fl\langle (N_1+1)(N_2-1) \rvert\, b_{r_2'}^\dagger b_{r_1'}^\dagger b_{r_1} b_{r_2} \,\lvert N_1 N_2 \rangle
\\
\lo=
N_1\sqrt{(N_1+1)N_2}\,\alphaindex{1}{2}{'}{*}\alphaindex{1}{1}{'}{*}
\bigl(\alphaindex{2}{1}{}{}\alphaindex{1}{2}{}{}+\alphaindex{1}{1}{}{}\alphaindex{2}{2}{}{}\bigr)
\\
+(N_2-1)\sqrt{(N_1+1)N_2}\,\bigl(\alphaindex{1}{2}{'}{*}\alphaindex{2}{1}{'}{*}+\alphaindex{2}{2}{'}{*}\alphaindex{1}{1}{'}{*}\bigr)
\alphaindex{2}{1}{}{}\alphaindex{2}{2}{}{}
\\
\fl
\langle (N_1+2)(N_2-2) \rvert\, b_{r_2'}^\dagger b_{r_1'}^\dagger b_{r_1} b_{r_2} \,\lvert N_1 N_2 \rangle
=
\sqrt{\fallingsmash{(N_1+2)}{2}\fallingsmash{N_2}{2}}\,
\alphaindex{1}{2}{'}{*}\alphaindex{1}{1}{'}{*}\alphaindex{2}{1}{}{}\alphaindex{2}{2}{}{}
,
}
\end{equation}
with all others zero.
For an expectation value (all $N_s'\<=N_s$), equation~(\ref{eqnmemulti}) simplifies to
\begin{equation}
\label{eqncoherentevmulti}
\eqalign{
\fl
\langle N_1\cdots N_S\rvert\, \Bigl(\prod_{i=1}^m b_{r_i'}^\dagger\Bigr)
\Bigl(\prod_{i=1}^m b_{r_i}\Bigr)|N_1\cdots N_S\rangle
\\
\lo=
\sum_{{t_1',\ldots,t_m' \atop t_1,\ldots,t_m} =1}^S 
\Bigl[\prod_{s=1}^S
\delta_{\nu_s',\nu_s}
N_s^{\underline{\nu_s}} \Bigr]
\Bigl( \prod_{i=1}^m \alpha_{t_i',r_i'}^* \alpha_{t_i,r_i} \Bigr) .
} 
\end{equation}

\section*{References}
\newcommand{\identity}[1]{{#1}}
\providecommand{\newblock}{}

\end{document}